\newcommand*{\RELEASE}{}  
\newcommand*{\NOCOPYRIGHT}{}  

\documentclass[conference]{IEEEtran}
\IEEEoverridecommandlockouts
\usepackage{cite}
\usepackage{amsmath,amssymb,amsfonts}
\usepackage{algorithmicx}
\usepackage{graphicx}
\usepackage{textcomp}
\def\BibTeX{{\rm B\kern-.05em{\sc i\kern-.025em b}\kern-.08em
    T\kern-.1667em\lower.7ex\hbox{E}\kern-.125emX}}

\graphicspath{{fig/}{../fig/}}

\usepackage[utf8]{inputenc}
\usepackage{paralist}  
\usepackage{balance}
\usepackage{xspace}  
\usepackage{soul}  
\usepackage{setspace}  
\usepackage{url}
\usepackage[bottom]{footmisc}  

\usepackage{booktabs}  
\usepackage{multirow}  
\usepackage{stfloats}  

\usepackage{subfigure}  
\usepackage{wrapfig}  
\usepackage{floatflt}  

\usepackage{algorithm}
\usepackage{algpseudocode}  

\usepackage[table]{xcolor}  

\definecolor{green(pigment)}{rgb}{0.0, 0.65, 0.31}

\ifdefined\RELEASE  
	\newcommand{\al}[1]{} 
	
	\newcommand{\del}[1]{}  

	\newcommand{\old}[1]{}  
	\newcommand{\objective}[1]{}  

	\newcommand{\dq}[1]{}
	\newcommand{\jgrm}[1]{}

	\newcommand{\pcm}[1]{}
\else
	\newcommand{\al}[1]{\textcolor{green(pigment)}{[AL: #1]}} 
	
	\newcommand{\del}[1]{\textcolor{blue}{\st{#1}}}  

	\newcommand{\old}[1]{\textcolor{gray}{#1}}  
	\newcommand{\objective}[1]{\textcolor{cyan}{[Objective: #1]\\}}  

	\newcommand{\dq}[1]{\textcolor{brown}{(*** DQ: #1 ***)}}
	\newcommand{\jgrm}[1]{\textcolor{brown}{\sout{#1}}}

	\newcommand{\pcm}[1]{\textcolor{red}{[PCM: #1]}}
\fi

\newcommand{\anm}{ANONYMIZED\xspace}
\ifdefined\ANONYMOUS
	\newcommand{\urlSys}{\anm} 
	\newcommand{\urlDaoc}{\anm} 
	\newcommand{\urlGembev}{\anm}  
\else
	\newcommand{\urlSys}{https://github.com/eXascaleInfolab/daor}
	\newcommand{\urlDaoc}{https://github.com/eXascaleInfolab/daoc}
	\newcommand{\urlGembev}{\url{https://github.com/eXascaleInfolab/GraphEmbEval}\xspace}  
	
\fi

\newcommand{\sys}{DAOR\xspace}

\newcommand\blfootnote[1]{%
  \begingroup
  \renewcommand\thefootnote{}\footnote{#1}%
  \addtocounter{footnote}{-1}%
  \endgroup
}

\begin{document}

\setstretch{0.958}  

\title{Bridging the Gap between Community\\ and Node Representations:\\ Graph Embedding via Community Detection
\thanks{This project has received funding from the European Research Council (ERC) under the European Union's Horizon 2020 research and innovation program (grant agreement 683253/GraphInt).}
}

\ifdefined\ANONYMOUS
\author{\IEEEauthorblockN{\anm}
\\
\\
}
\else
\author{\IEEEauthorblockN{Artem Lutov}  
\IEEEauthorblockA{\textit{eXascale Infolab}\\
\textit{University of Fribourg}\\
Switzerland\\
artem.lutov@unifr.ch}
\and
\IEEEauthorblockN{Dingqi Yang}  
\IEEEauthorblockA{\textit{eXascale Infolab}\\
\textit{University of Fribourg}\\
Switzerland\\
dingqi.yang@unifr.ch}
\and
\IEEEauthorblockN{Philippe Cudr{\'e}-Mauroux}  
\IEEEauthorblockA{\textit{eXascale Infolab}\\
\textit{University of Fribourg}\\
Switzerland\\
pcm@unifr.ch}
}
\fi

\maketitle

\begin{abstract}
Graph embedding has become a key component of many data mining 
and analysis systems.
Current graph embedding approaches 
either sample a large number of node pairs from a graph to learn node embeddings via stochastic optimization or factorize a high-order node proximity/adjacency matrix via computationally intensive matrix factorization techniques.
These approaches 
typically require significant resources for
the learning process and rely on multiple parameters, which limits their applicability in practice.
Moreover, most of the existing graph embedding techniques operate effectively in one specific metric space only (e.g., the one produced with cosine similarity), do not preserve higher-order structural features 
of the input graph and cannot automatically determine a meaningful number of dimensions for the embedding space. 
Typically, the produced embeddings are not easily interpretable, which complicates further analyses and limits their applicability.
To address these issues, we propose \sys, a highly efficient and parameter-free graph embedding technique producing metric space-robust, compact and interpretable embeddings without any manual tuning. Compared to a dozen state-of-the-art graph embedding algorithms, \sys yields competitive results on both node classification (which benefits form high-order proximity) and link prediction (which relies on low-order proximity mostly). Unlike existing techniques, however, \sys does not require any parameter tuning 
and improves the embeddings generation speed by several orders of magnitude.
Our approach has hence the ambition to greatly simplify and speed up data analysis tasks involving graph representation learning.
\end{abstract}

\begin{IEEEkeywords}
parameter-free graph embedding, unsupervised learning of network representation, automatic feature extraction, interpretable embeddings, scalable graph embedding.  
\end{IEEEkeywords}

\ifdefined\NOCOPYRIGHT
\else
	\blfootnote{978-1-7281-0858-2/19/\$31.00 \copyright~2019 IEEE}
\fi

\section{Introduction}
\label{sec:intro}

Representation learning has become a key paradigm to learn low-dimensional node representations from graphs. These automatically generated representations can be used as features to facilitate downstream graph analysis tasks such as node classification and link prediction. 
The main idea behind graph embedding techniques is to project graph nodes onto a low-dimensional vector space such that the key structural properties of the graph are preserved. The most commonly preserved property in this context is the \emph{proximity} between nodes in the graph~\cite{cai2018comprehensive}. For example, DeepWalk~\cite{Dpwk14} and Node2vec~\cite{nd2v16} preserve up-to-$k$-order node proximity by sampling random walk sequences from an input graph using a context window of a certain size; HOPE~\cite{Ou16} and NetMF~\cite{qiu2018network} capture high (up-to-infinite)-order node proximity by factorizing high-order node proximity matrices, measured by the Katz index~\cite{Katz53}, for example. The resulting embedding vectors from those techniques, capturing node proximity in graphs, are known to achieve good results on many downstream graph analysis tasks, such as node classification, node clustering (a.k.a. community detection) 
and link prediction.

Among various graph analysis applications, community detection is one of the most popular tasks.
Network communities (i.e., node clusters or granules~\cite{YaoY00})
represent groups of nodes that are
densely connected inside each group and
loosely connected between different groups~\cite{Newm03}.  
In essence, community detection techniques intrinsically capture node proximity in the graph to generate such node clusters.
In the context of graph embeddings, 
community detection naturally implies that nodes in a cluster should be projected closer to each other than to the nodes from other clusters in the vector space. For example, DeepWalk~\cite{Dpwk14} performs node sampling using random walks, such that nodes from the same cluster (intra-cluster nodes) are linked tighter together than nodes from different clusters (inter-cluster nodes) and have a higher probability to be closer in random walk sequences.
In the case of HOPE~\cite{Ou16}, the node proximity is defined by the Katz index~\cite{Katz53} that computes the weighted sum of all paths between two nodes. There, intra-cluster nodes also have a higher proximity measure than inter-cluster nodes, 
since there are more paths linking nodes in the same cluster than paths linking nodes between different clusters.

In this paper, by revisiting existing graph embedding techniques, we raise the following question: \emph{``Can we generate node embeddings from clusters 
produced via community detection?} 
More precisely, we explore how to generate node embeddings in a graph leveraging the latest advances in community detection techniques, analyzing and addressing emerging issues in the embedding learning process.

In the current literature, the graph embedding problem has also been investigated to specifically preserve community structures in a graph by applying hybrid approaches~\cite{ComE17,Wang17}. These approaches  
consist in specific graph embedding models jointly learning node embeddings and performing community detection,
where the two steps are performed iteratively until convergence. 
However, such hybrid methods preserving community structures lose other inherent advantages from modern community detection techniques:
%
\\$\bullet$ \emph{\textbf{Parameter-free}} community detection does not require any manual tuning, while current embedding techniques (hybrid or not) impose significant human efforts.
Parameter-free processing could significantly simplify the application of graph embeddings in practice compared to current graph embedding techniques, which  require manual tuning of multiple parameters (including the number of embedding dimensions).
\\$\bullet$ \emph{\textbf{Metric-Robustness}}: community detection typically does not rely on any specific metric space (e.g., cosine, Jaccard, or Hamming spaces), which provides an opportunity to obtain metric-robust node embeddings. More precisely, as existing graph embedding techniques are designed to learn node embeddings in a specific metric space (e.g., cosine \cite{Dpwk14} or Hamming \cite{yang2017histosketch}), they are often limited to the specified metric space. To ensure the applicability of learned embeddings in a wide variety of settings, it might hence be beneficial to learn embeddings that are robust to different metric spaces.
\\$\bullet$ \emph{\textbf{Efficient}} community detection techniques usually have linear or near-linear runtime complexity and are able to handle large graphs consisting of billions of nodes~\cite{Bld08}, which may significantly speedup the embedding learning process. Specifically, existing graph embedding techniques usually either sample a large number of node pairs from a graph to learn node embeddings via stochastic optimization, or factorize a high-order proximity/adjacency matrix, 
which requires significant computational resources. 
Therefore, it might be desirable to let graph embedding techniques benefit from the high-efficiency of state-of-the-art community detection techniques.

In this paper, we bridge the gap between community detection 
and node embeddings. More specifically, our main contributions can be formulated as follows:
\begin{inparaenum}[\itshape a\upshape)]
\item we propose a mechanism to generate node embeddings from given community structures (i.e., a set of clusters) and 
\item we identify the key features of community detection algorithms required to efficiently produce effective graph embeddings.
\end{inparaenum}
Both of these contributions are applied on top of the Louvain~\cite{Bld08}-based DAOC\footnote{\urlDaoc\label{ftn:daoc}}~\cite{Daoc19} clustering algorithm
and constitute  
our novel graph embedding framework called \sys\footnote{\urlSys\label{ftn:sys}}.

%
%
%
%
%
%
%

\section{Related Work}
\label{sec:relwork}

\subsection{Graph embedding}

%
Graph embedding techniques project graph nodes onto a low-dimensional vector space such that the key structural properties of the graph are preserved~\cite{cai2018comprehensive}. 
Existing techniques can be classified into four categories.
First, graph-sampling based techniques design specific embedding models to learn node embeddings from sampled node pairs from an input graph. The node pairs are often sampled by scanning random walk sequences from an input graph using a context window of size $k$ to capture up-to-$k$-order node proximity~\cite{Dpwk14,yang2019revisiting,hussein2018meta}, or directly sampled to capture 1st- and 2nd-order node proximities~\cite{tang2015line}.
Second, factorization-based techniques decompose specific node proximity matrices, such as high-order transitional matrices~\cite{cao2015grarep}, high-order proximity matrices measured by the Katz index, personalized PageRank or Adamic-Adar~\cite{Ou16}, to output node embeddings.
Third, hashing-based techniques resort to similarity-preserving hashing techniques~\cite{wu2018efficient,yang2019nodesketch} to create node embeddings, capturing high-order common neighbors between nodes in a graph. Due to the high efficiency of the hashing process, these techniques show significant speedup in the embedding learning process compared to the techniques of the two previous categories~\cite{yang2019nodesketch}.
Fourth, a few meta techniques are designed to preserve higher-order structural features by hierarchical coarsening of the input graph prior to the embedding process~\cite{Harp18}. 
As they capture higher-order node proximity in the graph, modern graph embedding techniques have shown good performance on various graph analysis tasks, including node classification, node clustering and link prediction.

In the current literature, the graph embedding problem has also been investigated to specifically preserve community structures in a graph~\cite{ComE17,Wang17}. These hybrid techniques combine the objectives of both node embedding learning and community detection. More precisely, community detection and node embedding learning are tightly coupled and are performed alternatively to enhance 
each other in the learning process.
The resulting node embeddings preserve the community structures of the input graph.
However, such hybrid approaches are not able to directly reuse results of existing community detection techniques, loosing their advantages as outlined in the introduction, i.e., they typically are not parameter-free, metric-robust, or particularly efficient (as we show in Section~\ref{sec:evals}).

In this paper, instead of jointly learning node embedding and detecting communities in a graph, we take an alternative solution to bridge the gap from detected communities to node embeddings.
Specifically, we
\begin{inparaenum}[\itshape a\upshape)]
\item design a new mechanism to generate node embeddings directly from given community structures (i.e., clusters) and
\item analyze the various aspects of community detection techniques required for effective node embeddings generation from the clusters.
\end{inparaenum}
	We implement these contributions in a novel graph embedding framework called \sys\hspace{-4pt}\footref{ftn:sys}. 

\subsection{Community detection for graph embedding}

To the best our knowledge, only matrix factorization-based (MF) community detection methods have been used to directly generate graph embeddings.
Spectral clustering is a matrix factorization approach~\cite{nd2v16}, which was the first community detection method used for graph embedding. It can be applied in two variations:
\begin{inparaenum}[\itshape a\upshape)]
\item conventional spectral clustering~\cite{Tang11} (introduced in~\cite{Fied73,Poth90} and operating on a Laplacian matrix) and 
\item spectral optimization of modularity~\cite{Tang09,Tang11} (introduced in~\cite{Newm06} and operating on a modularity matrix).
\end{inparaenum}
Node embeddings in this context are represented as 
the top-$d$ eigenvectors (i.e., latent dimensions) of the respective matrix. 
%
Conventional spectral clustering is equivalent to nonnegative matrix factorization (NMF)~\cite{Ding05}. The latter
is another community detection method~\cite{Yng13}, which is applied jointly with spectral optimization
of modularity  
to learn node embeddings~\cite{Wang17}.

The Gausian mixture model (GMM) is a statistical inference-based community detection method~\cite{Zhan07}, which can be used jointly with a conventional node representation learning (e.g., Deepwalk~\cite{Dpwk14} and  
SDNE~\cite{Sdne16}) to perform graph embedding~\cite{ComE17}. It is worth noting that GMM by itself explicitly learns the ``random mixtures over latent communities variables''~\cite{Zhan07} (i.e., node embeddings) but suffers from a large number of parameters and does not take into account the low-order proximity of the nodes
when generating the graph embeddings.
%
%

In essence, community detection by modularity maximization~\cite{Newm04}, statistical inference~\cite{Snij97}, 
normalized-cut graph partitioning~\cite{YuLb10}
and spectral clustering are equivalent under certain conditions~\cite{Newm13}. According to~\cite{Newm16}, community detection through generalized modularity maximization is equivalent to the provably correct but computationally expensive maximum likelihood method applied to the degree-corrected stochastic block model~\cite{Karr11,Peix14}. The latter inspired us to develop a novel graph embedding framework, which is able to generate node embeddings directly form the detected communities, and to extend an efficient community detection method based on parameter-free optimization of generalized modularity to produce effective embeddings.

%
%
%
%

\section{Preliminaries}  
\label{sec:prelim}

Network communities 
represent groups of tightly-coupled graph nodes with loosely inter-group connections~\cite{Newm03}, where the group structure is determined by a clustering optimization function.
The resulting clusters can be overlapping, 
which happens in case they share some common nodes called the overlap\index{overlap}. Also, the clusters can be nested, forming a hierarchical structure inherent to many complex real-world systems~\cite{HSim62}. Each cluster represents a coarse-grained view on (i.e., an approximation of) its member nodes or subclusters being called a granule~\cite{YaoY01}.  
%
The main notations used in 
this paper are listed in Table~\ref{tbl:nots}.
\begin{table}[htbp]\small  
\centering
\caption{Notations}
\label{tbl:nots}
\vspace{-4pt}
\rowcolors{1}{gray!25}{white}
\begin{tabular}{l|l}
\hline
{\it \verb|#i|}   	& {Node \verb|<i>| of the graph (network) $\mathcal{G}$}
\\ 
$c_i$				& {Cluster \verb|<i>| of the graph $\mathcal{G}$}
\\ 
$n$				& {The number of nodes in the graph $\mathcal{G}$}
\\ 
$m$				& {The number of links in the graph $\mathcal{G}$}
\\ 
$k$				& The number of clusters (communities) in the graph $\mathcal{G}$ 
\\ 
$w$				& {Weight of the graph $\mathcal{G}$}
\\ 
$w_i$				& Weight of {\it \verb|#i|} 
\\ 
$\dot{w}_i$				& Weight of $c_i$: $\dot{w}_i = w_{c_i}$ 
\\ 
\textit{Q}				& \emph{Modularity}
\\ 
$\Delta Q_{i,j}$	& \emph{Modularity Gain} between \verb|#i| and \verb|#j|
\\ 
\textit{$\gamma$}				& \emph{Resolution} parameter
\\ 
$s$				& The number of salient clusters (features), $s \le k$
\\ 
$d$				& The number of embedding dimensions: $d = |D|, d \le s$
\\ 
\hline
\end{tabular}
\vspace{-4pt}
\end{table}

\emph{Modularity} ($Q$)~\cite{Newm04u} is a standard measure of clustering quality 
that is equal to the difference between the density of the links in the clusters and their expected density: 
\begin{equation}  
Q = \frac{1}{2w}\sum_{i,j}\left({w_{i,j} - \frac{w_i w_j}{2 w}}\right)\delta(C_i, C_j),
\label{eq:mod}
\end{equation}
where $w_{i,j}$ is the accumulated weight of the links 
between nodes \verb|#i| and \verb|#j|; $w_i$ is the accumulated weight of all links 
of \verb|#i|; $w$ is the total weight 
of the graph; 
$C_i$ is the cluster to which \verb|#i| is assigned; and the Kronecker delta $\delta(C_i, C_j)$ equals to $1$ when \verb|#i| and \verb|#j| belong to the same cluster (i.e., $C_i = C_j$), and $0$ otherwise.

\emph{Modularity gain} ($\Delta Q$)~\cite{Bld08} captures the difference in modularity when merging two nodes \verb|#i| and \verb|#j| into the same cluster, providing a computationally efficient way to optimize Modularity:
\begin{equation}
\Delta Q_{i,j} = \frac{1}{2 w} \bigg( w_{i,j} - \frac{w_i w_j}{w} \bigg).
\label{eq:dmod}
\end{equation}

Modularity 
is commonly used as a global optimization criterion but suffers from the \emph{resolution limit}\index{modularity!resolution limit} problem~\cite{Fort07,Good10}, which corresponds to its inability to detect clusters smaller than a certain size. To address this problem,
\emph{generalized modularity} was proposed with a resolution parameter $\gamma$~\cite{Aren08r,Kump07}:  
\begin{equation}  
Q = \frac{1}{2 w}\sum_{i,j}\left({w_{i,j} - \gamma \frac{w_i w_j}{2 w}}\right)\delta(C_i, C_j).
\label{eq:gmod}
\end{equation}
The optimal value of the resolution parameter is $\acute{\gamma}$~\cite{Newm16}: 
\begin{align}\begin{split}
\acute{\gamma} = &\frac{\check{p} - \hat{p}}{\log \check{p} - \log \hat{p}},\\
\check{p} = \frac{2 \check{w}}{\sum_{c}\dot{w}^2_c / (2 w)},\quad
&\hat{p} = \frac{2 w - 2 \check{w}}{2 w - \sum_{c}\dot{w}^2_c / (2 w)},
\label{eq:gamma}
\end{split}\end{align}
where $\check{w}$ is the total internal weight of all clusters (i.e., accumulated weight of all intra-cluster links).
The generalized modularity is equivalent to the standard modularity when $\gamma = 1$; it tends to find larger (macro-scale) clusters if $\gamma \in [0, 1)$ and smaller clusters otherwise.
We use the \emph{generalized modularity gain} ($\Delta Q$) 
as an underlying optimization function for the meta-optimization strategy MMG of the DAOC~\cite{Daoc19} clustering algorithm on top of which our framework, \sys, is built.



\section{Transforming Clusters into Node Embeddings}
\label{sec:dims}

\begin{figure}[htbp]\centering  
\includegraphics[scale=0.5]{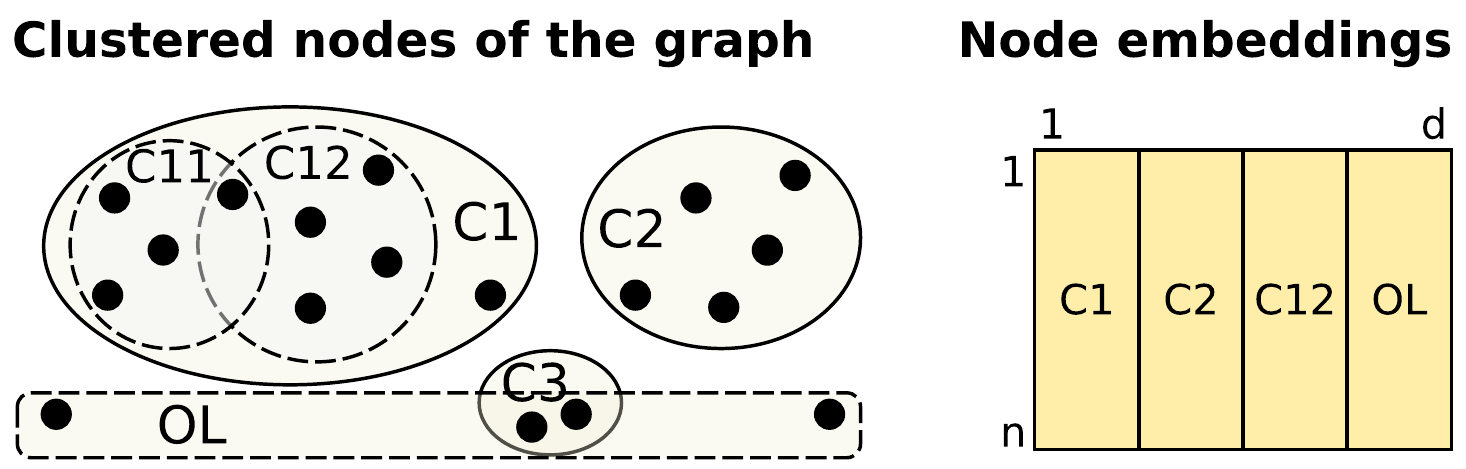}
\caption{Transformation of the hierarchy of overlapping clusters consisting of $n$ nodes and $k=5+2$ clusters 
(2 nodes-outliers, which are grouped as OL together with the outlier cluster C3) into $d=4$ dimensional node embeddings.
}
\label{fig:clsemb}
\end{figure}

Community detection algorithms only generate clusters as groups of nodes, which hence requires some post-processing
to produce node embeddings, namely to:
\begin{inparaenum}[\itshape a\upshape)]
\item form latent embedding dimensions from the clusters (i.e., extract features) and
\item quantify node membership in each dimension (i.e., generate the embedding vector for a node).
\end{inparaenum}
In addition, it is often desirable to manually control the number of embedding dimensions $d$.
These aspects are described in the following and are illustrated in Fig.~\ref{fig:clsemb}.

\subsection{Feature Extraction from Clusters} 
\label{subsec:features}

Intuitively, a straightforward approach for feature extraction from clusters is to consider each cluster as a dedicated feature representing one embedding dimension~\cite{Tang09}. This approach can be used to discriminate between the nodes (given a fine-grained structure of clusters), but on the other hand, may not correlate with the graph embedding objective of providing a \emph{low-dimensional} representation of the nodes. Fixing the number of clusters to the required number of dimensions $d$ is possible when considering flat clustering~\cite{Tang09} but not for hierarchical cases. For hierarchical clustering, producing $d$ dimensions can be expressed with an inequality in the general case, i.e. producing \emph{at most $d$} clusters at the top level.
Therefore, some technique to identify a small number $s \ge d$ of 
\emph{salient} clusters, which are then regarded as features, 
is required.  
This 
number $s$ can either be parameterized or, ideally, inferred as the optimal tradeoff between the number of clusters and their quality. 
%
In addition, there exists a fundamental constraint on the structure of the clusters used to create the embedding space. Namely, each graph node should be connected to at least one feature to be present in the embeddings.


The exact way salient clusters (features) are identified depends on the structure of the clusters (e.g., hierarchical, multi-resolution, overlapping, non-overlapping) and on the ability to control the number of formed clusters $k \ge s$ by the community detection algorithm.
In the following, we discuss salient clusters identification for the most general case, i.e., for clustering techniques that are multi-resolution, hierarchical and overlapping. 
\begin{algorithm}[htbp]\small 
\caption{Features extraction}
\label{alg:features}
\begin{algorithmic}[1]  
\Function{features}{$hier$}
	\State $scls \gets []$  \Comment Salient clusters (features), dynamic array
	\State $clsts \gets \{\}$  \Comment Ancestor cluster statistics, hash map
	\ForAll{$lev \in hier$}
		\ForAll{$cl \in lev$}
			\State $res \gets \texttt{count}(cl.ances) = 0$  \label{aln:saltop}  \Comment Salient flag
			\State $dens \gets cl.weight / \texttt{count}(cl.nodes)$  \Comment Density
			\State $savdens \gets 0; savwgh \gets 0$  \Comment Statistics to be stored
			\If{\textbf{not} $res$}
				\State $hits \gets 0$  \Comment Saliency hits
				\ForAll{$ac \in cl.ances$}  \Comment Traverse ancestors
					\State $ast \gets clsts[ac]$  \Comment Ancestor statistics
					\If{$savdens < ast.dens$ \textbf{and} (\textbf{not} $savwgh$ \textbf{or} $savwgh > ast.wgh$)}
						\State $savdens \gets ast.dens$
						\State $savwgh \gets ast.wgh$
					\EndIf
					\If{\textbf{not} $res$ \textbf{and} $dens \ge ast.dens$ \textbf{and} $cl.weight \le ast.wgh$} \label{aln:saldes}
						\State $hits \gets hits + 1$
					\EndIf
					\State $ast.reqs \gets ast.reqs + 1$
					\If{$ast.reqs = \texttt{count}(ac.des)$}  
						\State $clsts.erase(cl)$  \Comment Remove outdated
					\EndIf
				\EndFor
			\EndIf

			\If{\textbf{not} $res$ \textbf{and} $hits = \textbf{count}( cl.ances)$}
				\State $res \gets$ true
			\EndIf

			\State $savwgh \gets cl.weight * wrstep$ \label{aln:wrmul}
			\State $clsts[cl] \gets (savdens, savwgh)$  \Comment dens,wgh,reqs=0
			\If{$res$}
				\State $scls.add(cl)$
			\EndIf
		\EndFor
	\EndFor

	\State \Return $scls$  \Comment{The resulting salient clusters}
\EndFunction
\end{algorithmic}
\end{algorithm}
Features extraction is presented in Algorithm~\ref{alg:features}.
We traverse all clusters on each level of the formed hierarchy of clusters starting from the top level, and fetch as salient clusters:
\begin{inparaenum}[\itshape a\upshape)]
\item the top-level clusters of the hierarchy (line~\ref{aln:saltop}) (to cover all nodes) and
\item all the nested clusters satisfying the following requirements (line~\ref{aln:saldes}):
\end{inparaenum}
\\- having a higher weight density 
than each of their direct super-clusters (ancestors), 
since the nested salient clusters 
are expected to represent the mutual core structure~\cite{Borg00} 
of their ancestors, and
\\- weighting less than 
the most lightweight ancestor (discounted by a factor).
\\The weight discounting factor $r_w$ (line~\ref{aln:wrmul}) 
is required to prevent fetching too many similar clusters that are nested into each other. 
The factor $r_w \in [0.5, 1)$ retains an approximation of the nested cluster to its ancestor, while $r_w \rightarrow r_{w_{min}} = 0.5$ reduces the number of salient clusters. However, considering the availability of overlapping clusters and known weight of the overlap $w_{C_{ovp}} < w_C$ 
for the cluster $C$, we refine $r_w$ as $r_w = 0.5 + \frac{(b - 1) \times w_{c_{ovp}}}{2\, b \times w_c}$, where $b \ge 2$ is the overlap factor equal to the number of clusters sharing the overlapping nodes in $C$.
Also, the number of overlapping clusters can be estimated according to the Pareto principle~\cite{Pareto,Geor86}, which makes reasonable to take $r_{w} = 0.5 \times 1.2 = 0.6$ when the exact value cannot be evaluated. 


\subsection{Dimension Formation from Features}  
\label{subsec:dimform}

Embedding dimensions are formed from the salient clusters extracted above, which implicitly yields the recommended number of dimensions for the input graph based on its statistical properties.
The embedding vector $v_i \in V$ of size $d = |D|$ for each graph node \verb|#i| is then produced by quantifying the degree of belonging of the node to each dimension $D_j$ as the node weight $w_{i,D_j}$ belonging to this dimension. This weight corresponds to the aggregated weight of the node links to other nodes being members of all salient clusters composing the dimension:
\begin{align}\begin{split}
V = \{ \frac{w_{i,D_j}}{w_i}\; |\; i = 1..n, j = 1..d \},\\
w_{i,D_j} = \sum_{k \in nodes(D_j)} w_{i, k}\,.
\end{split}\end{align}
We perform this embedding vectors generation efficiently by resorting to a single scan over the members of each salient cluster. The node weights are then 
aggregated to form the dimension values. The embedding vectors are obtained by transposing the resulting matrix: $V = D^T$. 

\paragraph{Constraining the Number of Dimensions}
In the case of a connected graph without clusters-outliers, the most salient clusters $d$ are fetched from the $t \le d \le s$ top level clusters of the hierarchy and from the $d - t$ densest of the remaining $s - t$ salient clusters. 
When the clustering algorithm does not allow to control the number of top level clusters or when the graph is disconnected and the number of components is larger than $d$ resulting in $t > d$, then the dimensions are formed as follows. According to the so-called ``Rag Bag'' formal constraint of clustering quality~\cite{Xms19,Amg09}, the $t - (d-1)$ most lightweight clusters should be grouped together to the last dimension and the $d - 1$ heaviest clusters fill the remaining dimensions.
However, the presence of outliers on the top levels prevents to effectively generate dimensions from the salient clusters. To solve this issue, the outliers can be separated from the valid clusters
based on their weights, which are either evaluated from the statistical distributions or approximately estimated as follows.
In case there is no prior information about the data, a rule of thumb is to take the estimated minimal size of clusters as the square root of the number of nodes~\cite{Mard79}. Generalizing the rule of thumb to the weighted graphs, the number of $z < t$ root clusters, each having a weight less that the square root of the graph weight $w$ can be moved to the ``outliers'' dimension. The resulting dimensions are composed of the $t - z$ top level clusters of the hierarchy,
the $d - 1 - (t - z)$ densest remaining salient clusters, each having a weight $w_i \ge \sqrt{w}$, and a single dimension of outliers to cover all remaining graph nodes (see Fig.~\ref{fig:clsemb} for an illustration).  

\paragraph{Dimension Interpretability}
The resulting embedding space is inherently easier to interpret than spaces derived from other techniques, as its dimensions are taken from (salient) clusters representing ground-truth 
semantic categories, 
with accuracy being evaluated using extrinsic quality metrics~\cite{Xms19}.  
This opens the door to new features and applications in the fields of granular and privacy-preserving computations. For example, only a subset of the dimensions having some required semantics can be fetched for evaluation, which has the advantage of reducing the amount of processing data while avoiding to leak or share information beyond the required features.



\section{Community Detection}
\label{sec:calg}

In this section, we first discuss the properties 
of a community detection algorithm that are required to perform an effective and efficient graph embedding. 
Then, we select one of the most suitable state-of-the-art community detection algorithms for our task and propose its extension to satisfy the required properties.

As an effective graph embedding technique should preserve both low- and high-order node proximities, the community detection algorithm used for the graph embedding should be able to produce clusters with various resolutions (i.e., at different granularities). Moreover, the more resolutions are covered, the wider the range of node proximity orders that can be captured, since each embedding dimension consists of at least one cluster as described in Section~\ref{subsec:dimform}.
%
A low-dimensional representation of graph nodes 
implies a \emph{small} number $t$ of coarse-grained (macro-scale) clusters, since the number of generated dimensions $d \ge t$ (see Section~\ref{sec:dims}). This number $t$ should be a parameter of the technique when a specific number $d$ of embedding dimensions is required, with a default value defined according to the statistical properties of the input graph.

In addition, the following properties of community detection algorithms are also required to generate high-quality embeddings, to simplify and speedup the generation process:
\begin{itemize}
\item Each graph node should potentially belong to multiple features (i.e., should be represented in multiple dimensions), which requires the clustering to be soft (i.e., fuzzy or \emph{overlapping}).  
\item It is desirable to have graph embedding techniques applicable to any  input graph without any manual tuning; hence, the clustering should be \emph{parameter-free} and \emph{efficient} (to be applicable to large graphs).
\\- An unsupervised parameter-free processing is sensitive to the quality of the input data, so a \emph{robust} clustering algorithm is required. Robustness is typically reached by applying a consensus (also called ensemble) clustering method~\cite{Frd03,VgP11,Lnc12}.
\\- From a practical perspective, it is desirable to have consistent embeddings for the same input graph irrespective of the order in which the nodes are processed or whether their IDs are modified.
The clustering algorithm should hence be \emph{deterministic} and input-order invariant.
\\- Considering the hierarchical nature of complex networks modeling real-world systems~\cite{HSim62,Bara16}, the effective clustering algorithm should be hierarchical. 
In particular, an agglomerative hierarchical clustering addresses also the efficiency criterion by reducing the number of processed items at each iteration, since each hierarchy level is built using clusters from the previous level directly. 
\end{itemize}

Following the above requirements, DAOC\footref{ftn:daoc}
~\cite{Daoc19} is, to the best of our knowledge, the only parameter-free clustering algorithm that is simultaneously deterministic, input order invariant, robust (as it uses a consensus approach) and applicable to large weighted networks yielding a fine-grained hierarchy of overlapping clusters~\cite{Clb18}. Moreover, it is based on a MMG meta-optimization function, where generalized modularity gain can be used as the target optimization function to perform clustering at the required resolution.
However, DAOC
\begin{inparaenum}[\itshape a\upshape)]
\item yields a hierarchy of clusters only for a single value of the resolution parameter ($\gamma = 1$ operating with the non-generalized modularity),
treating the hierarchy levels as scales (i.e., resolutions in terms of nested clusters rather than distinct values of $\gamma$) similar to~\cite{SalP07,Bld08}, 
and
\item does not bound the size of the top (root) level of the forming hierarchy to produce the required number $t \le d$ of clusters as described in Section~\ref{subsec:features}.
\end{inparaenum}
Therefore, we propose two extensions addressing these issues in the following.

\subsection{Hierarchical multi-resolution clustering}

\begin{figure}[htbp]\centering  
\includegraphics[scale=0.20]{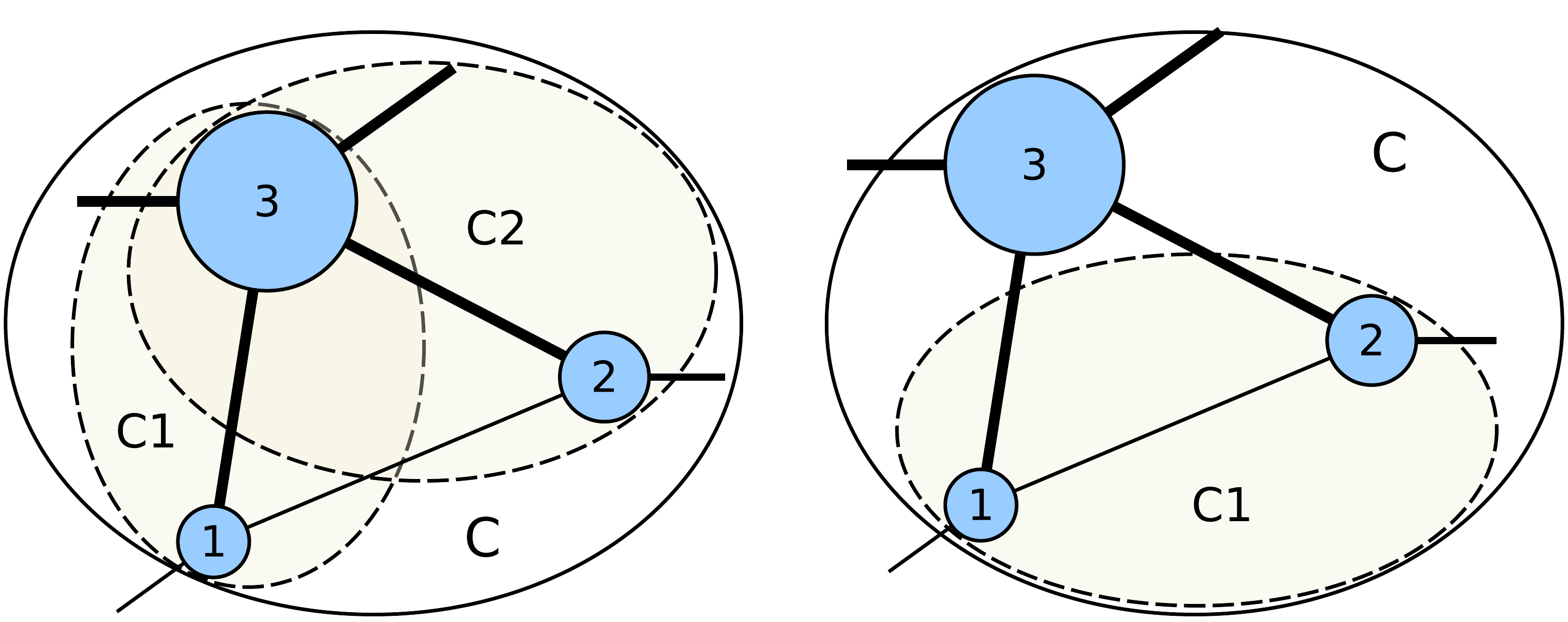}
\caption{A possible overlapping hierarchical clustering with the fixed resolution parameter of the weighed subgraph consisting of three nodes is shown on the left-hand side of the figure. A possible multi-resolution clustering for the same subgraph is shown on the right-hand side. 
The size of the nodes and the width of the links correspond to their weights.
}
\label{fig:hovp_mres}
\end{figure}

Even though a multi-resolution structure of non-overlapping clusters is not necessary hierarchical~\cite{Aren08,Lamb10} (i.e., a node might be a member of several clusters that are not nested into each other), it can be represented as a \emph{hierarchy of overlapping clusters}~\cite{Lanc09} (where the overlap addresses the case of a node shared by non-nested clusters). However, a hierarchical overlapping structure created with a single resolution may substantially differ from the respective multi-resolution structure (i.e., the produced clusters may differ vastly) as illustrated in Fig.~\ref{fig:hovp_mres}, where the strength on nodes interaction is represented with the width of the links and the number of interactions with the size of the nodes (i.e., their weight). A large value of the resolution parameter can penalize heavily-weighted nodes according to Eq.~\eqref{eq:gmod}, resulting in grouping together linked 
lightweight nodes. 
Such a behavior makes sense in many real-world cases, for example when employees working on the same project interact more frequently with their supervisor than between each other but the supervisor may not be a core of the group, participating also in other projects.
Therefore, it is essential to incorporate the resolution parameter when generating the hierarchy levels, similar to~\cite{Aren08,Pons11,Gran12}.
%

In~\cite{Aren08,Gran12}, the authors operate with the analog of the resolution parameter called resistance parameter, which can not be transformed to the resolution parameter $\gamma$ according to~\cite{Aren08r} and, hence, cannot be used with the generalized modularity defined in Eq.~\eqref{eq:gmod}. In~\cite{Pons11}, the scale factor $\alpha$ is proposed, which can be directly transformed to the resolution parameter: $\gamma = (1 - \alpha)/\alpha$. However, the computational complexity of the proposed method is $O(n\sqrt{n}) \times O_0$ in average and $O(n^2) \times O_0$ in the worst case, where $n$ is the number of nodes in the graph and $O_0$ is the complexity of the original clustering algorithm without the multi-scaling technique. This boosting of the computational complexity makes this technique inappropriate for large graphs. Hence, we propose our own approach to produce a hierarchy of multi-resolution clusters for large graphs based on DAOC.

The main idea behind our efficient multi-resolution hierarchical clustering is the dynamic variation of the resolution parameter $\gamma$ during the hierarchy construction process. More precisely, DAOC is an agglomerative clustering algorithm, which builds the hierarchy bottom-up
with micro-scale clusters on the bottom levels. These clusters should be constructed on a high value of the resolution parameter ($\gamma \gg 1$) according to Eq.~\eqref{eq:gmod}. 
The resolution should then gradually decrease to the lowest value $\gamma > 0$ yielding macro-scale clusters 
located on the top (root) level of the hierarchy.
The bounds for $\gamma$ can be estimated based on the resolution limit analysis conducted in~\cite{Fort07}, where the equation is defined
for the marginal case when all clusters have a perfect balance between the internal and external degrees being on 
the limit of detectability. 
That equation takes the following shape when adapted to the 
generalized modularity:
\begin{equation}  
\check{\dot{m}} < \check{\dot{m}}_{max} = \frac{m}{4 \gamma},
\label{eq:cllns}
\end{equation}
where $\check{\dot{m}}$ is the number of internal links in a cluster and $m$ is the total number of links in the graph. To estimate the upper bound of $\gamma$, we need to bind $\check{\dot{m}}$ and $m$, which could be done relying on the two following heuristics. First, in case there is no prior information about the data, a rule of thumb is to take the estimated maximal number of clusters as the square root of the number of nodes~\cite{Mard79}: $\tilde{k}_{max} = \sqrt{n}$. Then, considering that the number of internal links constitutes half of all the links of a cluster in the marginal case described by Eq.~\eqref{eq:cllns}:
\begin{equation}
2\,m = 2\,\check{\dot{m}} \times k,\; k \lesssim \tilde{k}_{max} \implies m \lesssim \check{\dot{m}} \sqrt{n}.
\label{eq:grlns}
\end{equation}
Second, most real-world systems are modeled by sparse graphs~\cite{Bara16}. The number of links in a sparse graph does not exceed $n^{3/2}$. Thereby, Eq.~\eqref{eq:cllns} extended with Eq.~\eqref{eq:grlns} takes the following shape:
\begin{equation}  
\check{\dot{m}} < \frac{\check{\dot{m}} \sqrt{n}}{4 \gamma} \implies
\gamma < \frac{\sqrt{n}}{4} \lesssim \frac{\sqrt[3]{m}}{4} \le \frac{\sqrt[3]{w/\breve{w}_{min}}}{4} = \gamma_{max}.
\label{eq:gbup}
\end{equation}
where $\breve{w}_{min}$ is the minimal weight of a link. Eq.~\eqref{eq:gbup} 
provides a definition for the upper bound of $\gamma$ in a weighted graph.
The lower bound of gamma is evaluated dynamically based on the optimal value of the resolution parameter given in Eq.~\eqref{eq:gamma} for each level of the hierarchy: $\gamma_{min} = \acute{\gamma}$.
Typically, $\acute{\gamma} \ge 1$ for large real-world networks~\cite{Newm16}, so $\gamma_{min} = 0.5\, .. \,1$ is taken for the first iteration before $\acute{\gamma}$ is evaluated. 

In summary, $\gamma$ is decreased from $\gamma_{max}$ to $\gamma_{min}$ with a fixed ratio $r_\gamma$  
and represents a trade-off between the number of captured resolutions when $r_\gamma \rightarrow 1$ versus a lower number of iterations (i.e., higher efficiency) and higher quality on coarse resolutions when $r_\gamma \rightarrow 0$. The quality of the forming macro-clusters is affected by the amount and size of the fine-grained clusters since the former are constructed iteratively from the latter, 
and the actual multi-resolution structure is not necessary hierarchical~\cite{Aren08,Lamb10}.
In theory, $r_\gamma$ should allow the growth of the cluster weight by a factor less than two 
to still retain the cluster semantics (super/sub-cluster ordering). 
Otherwise, the super/sub-cluster association is transformed into an overlapping clusters relation, losing the hierarchical structure.
The limitation of the cluster weight growth to a factor 2 corresponds to $r_\gamma > 2^{-2} = 0.25$
according to Eq.~\eqref{eq:gmod}, which is a hard theoretical bound.
In practice,
a larger value of $r_\gamma$ is required considering the possibility of multiple mutually overlapping clusters as discussed in Section~\ref{subsec:features}:  $r_\gamma \gtrsim r_{w_{min}}^2 = 0.36$.
The upper bound of $r_\gamma$ corresponds to the lower bound of the cluster weight growth factor, which can be taken as $10-20\%$ according to the Pareto principle~\cite{Pareto,Geor86} or 10/90 gap~\cite{Curr18}: $r_\gamma \le 
1.1^{-2} = 0.826$. Thus, the operational range of the gamma ratio is as follows: $r_\gamma \in [0.36, 0.826]$, and we pick $r_\gamma = 0.6$.
This selected value 
is not supposed to be tuned by the users. Higher values of $r_\gamma$ 
yield a larger number of salient clusters, which is not desirable, and lower values may cause the loss of some important salient clusters. The exact theoretical $r_\gamma$ depends on the number of overlapping clusters on 
the next iteration, which is generally speaking unknown. 

\subsection{Bounding the number of clusters}
\label{subsec:clsbound}

DAOC generates a hierarchy of clusters as long as the value of the optimization function is improving (i.e., $\Delta Q \ge 0$), which might result in any number of clusters at the top level of the hierarchy.
If a specific number of clusters is required (e.g., for a fair comparison of various graph embedding techniques on the same number of dimensions), 
we propose the following extensions of the hierarchy generation process:
\begin{itemize}
\item The hierarchy generation is interrupted early if the number of clusters at level $i$, $|h_i|$ reaches 
the required number $d$.
\item The hierarchy generation is forced to continue until the number of clusters reaches 
the required number $d$ even if the value of the optimization function $\Delta Q$ becomes negative. In that case, the clustering objective becomes the minimal loss 
of the optimization function value: $ \max \Delta Q < 0$.
\end{itemize}

The proposed extensions of the clustering algorithm are
summarized in Algorithm~\ref{alg:daocx}. 
The early termination of the hierarchy construction process happens on lines~\ref{aln:cterm1},~\ref{aln:cterm2}. The forced continuation of the hierarchy construction corresponds to a boolean parameter $d \ne 0$ for the original DAOC clustering on line~\ref{aln:hierLev}. This parameter prevents the completion of the clustering process when the optimization function $\Delta Q$ cannot be further improved.
\begin{algorithm}[H]\small 
\caption{Hierarchical multi-resolution clustering with optionally bounded number of clusters.}
\label{alg:daocx}
\begin{algorithmic}[1]  
\Function{cluster}{$nodes, d$}
	\State $rg \gets 0.6$  \Comment Gamma ratio
	\State $gamma \gets \sqrt[3]{\texttt{weight}(nodes)/\texttt{min\_weightln}(nodes)}/4$  
	\State $gmin \gets 0.5$  \Comment Min gamma
	\State $hier \gets []$  \Comment List of the hierarchy levels 
	\While{$nodes$} \Comment Stop if the \textit{nodes} list is empty
		\State $cls \gets \texttt{daocHierLev}(nodes, gamma, d)$ \label{aln:hierLev}  

		\If{$gamma * (rg + 1) / 2 \ge gmin$ \textbf{and} (\textbf{not} $cls$ \textbf{or} $\texttt{count}(cls) \le \sqrt{\texttt{count}(nodes)}$)}  \Comment Decrease gamma  
			\State $gamma \gets gamma * rg$
		\Else
			\State $nodes \gets cls$ \label{aln:cterm1}  \Comment Consider early termination 
		\EndIf

		\If{$cls$}  \Comment Initialize the next-level nodes  
			\State $hier.append(cls)$  \Comment Extend the hierarchy  
			\State $gmin \gets \texttt{gammaOptim}(cls)$  \Comment Update min gamma
			\If{\textbf{not} $d$ or $\texttt{count}(cls) > d$}
				\State $nodes \gets cls$  \Comment Update the processing nodes 
			\Else
				\State $nodes \gets []$ \label{aln:cterm2}  \Comment Early termination
			\EndIf
		\EndIf

	\EndWhile
	\State \Return $hier$  \Comment{The resulting hierarchy of clusters}
\EndFunction
\end{algorithmic}
\end{algorithm}

\section{Experimental Evaluation}  
\label{sec:evals}

In this section, we conduct an extensive set of experiments to evaluate our proposed method on two typical graph analysis tasks, i.e., node classification and link prediction. We start by introducing our experimental setup including the datasets and baselines we use before presenting our experimental results.

%
%
%
%

\subsection{Experimental Setup}

\subsubsection{Datasets}
We conduct experiments on the following five graphs, which are widely used in the current literature for evaluating graph embedding techniques. Table \ref{dataset} summarizes the key characteristics of the graphs.
\\$\bullet$ \emph{BlogCatalog (\textbf{Blog})}~\cite{Tang09} is a social network of bloggers. Each node represents a user, while the labels of a node represent the 
topics the corresponding user is interested in.
\\$\bullet$ \emph{Protein-Protein Interactions (\textbf{PPI})}~\cite{nd2v16} is a graph of the PPI network for Homo Sapiens. The labels of a node refer to its gene sets and represent the corresponding biological states.
\\$\bullet$ \emph{Wikipedia (\textbf{Wiki})}~\cite{nd2v16} is a co-occurrence network of words appearing in a sampled set of the Wikipedia dump. The labels represent part-of-speech tags.
\\$\bullet$ \emph{\textbf{DBLP}}~\cite{yang2015defining} is a collaboration network capturing the co-authorship of writers. Each node represents an author, and the labels of a node refer to the publication venues of the corresponding author. 
\\$\bullet$ \emph{\textbf{YouTube}}~\cite{tang2009scalable} is a social network of users on YouTube. Each node represents a user, and the labels of a node refer to the groups (e.g., ``anime'') that the corresponding user is interested in. This graph is used only to evaluate the efficiency of the embedding techniques, 
since the ground-truth categories include only 
3\% of the graph (as opposed to a 100\% coverage for the other graphs). 
%
\begin{table}[htbp]\small
\vspace{-4pt}
\centering
\caption{Characteristics of the experimental graphs}
\label{dataset}
\vspace{-4pt}
\rowcolors{2}{gray!25}{white}
\begin{tabular}{l|rrrrr}\hline
\textbf{Dataset}		&\textbf{Blog}		&\textbf{PPI}	&\textbf{Wiki}	&\textbf{DBLP} 	&\textbf{YouTube}\\ \hline
Nodes			&10,312			&3,890 	&4,777	&13,326 		&1,138,499\\ 
Links			&333,983			&76,584 	&184,812	&34,281		&2,990,443\\ 
Labels			&39 				&50	 	&40		&2			&47\\ \hline
\end{tabular}
\end{table}

\subsubsection{Baselines}
We compare our proposed technique against ten state-of-the-art graph embedding techniques from three categories.
\paragraph{Graph-sampling based techniques} \textbf{DeepWalk}~\cite{Dpwk14}, \textbf{Node2Vec}~\cite{nd2v16}, \textbf{LINE}~\cite{tang2015line} and \textbf{VERSE}~\cite{tsitsulin2018verse}. For DeepWalk and Node2Vec, we set the walk length to 40, the number of walks per node to 80, and the context window size to 10. For Node2Vec, we also tune the return parameter $p$ and the in-out parameter $q$ with a grid search over $p,q \in \{0.25,0.05,1,2,4\}$. For LINE, we set the total number of samples to 1 billion for Blog, PPI, Wiki and DBLP and to 10 billions for YouTube. For VERSE, we tune the damping factor $\alpha$ of personalized PageRank using the method suggested by the authors.

\paragraph{Factorization-based techniques} \textbf{GraRep}~\cite{cao2015grarep}, \textbf{HOPE}~\cite{Ou16} and \textbf{NetMF}~\cite{qiu2018network}. For GraRep, we search the optimal $k$ over $\{1,2,3,4,5,6\}$. When $d/k$ is not an integer, we learn the first $k-1$ sets of $\lceil d/k \rceil$-dimension embeddings, and the last set of embeddings of dimension $d-(k-1)\lceil d/k \rceil$. For HOPE, we search the optimal decay parameter $\beta$ from $0.1$ to $0.9$ with a step of $0.2$. For NetMF, we tune the implicit window size $T$ within $\{1,10\}$.

\paragraph{Similarity-preserving hashing based techniques} \textbf{INH-MF}~\cite{lian2018high}, \textbf{NetHash}~\cite{wu2018efficient} and \textbf{NodeSketch}~\cite{yang2019nodesketch}. For INH-MF, we set the ratio for subspace learning to 100\%, to let it achieve optimal performance w.r.t. the quality of the learnt node embeddings. For NetHash, as suggested by the authors, we search the optimal tree depth in $\{1,2,3\}$. For NodeSketch, we search the optimal order of proximity $k$ up to $6$ and the optimal decay parameter $\alpha$ from $0.0001$ to $1$ on a log scale.

\paragraph{Meta techniques} \textbf{HARP}~\cite{Harp18}.
We configure HARP to learn from the embeddings of DeepWalk (HARP-DWalk) and LINE (HARP-LINE) using the following parameter settings. 
For HARP-DWalk, we set the walk length to 10, the number of walks per node to 40, the context window size to 10 and the sampling ratio to 0.1. For HARP-LINE, we set the context window size to 1, the number of iterations to 50 and the sampling ratio to 0.001.

\subsection{Node Classification Task}

Node classification tries to predict the most probable label(s) for some nodes based on other labeled nodes. In this experiment, we consider a multi-label setting, where a node is assigned one or multiple labels. Our evaluation was performed using an open-source graph embeddings evaluation framework, GraphEmbEval\footnote{\urlGembev\label{ftn:gembev}}, which uses the same evaluation scheme as in~\cite{Dpwk14,nd2v16,cao2015grarep}. More precisely, we randomly pick a set of nodes as labeled nodes for training, and use the rest for testing. To fairly compare node embeddings with different similarity measures, we train a one-vs-rest kernel SVM classifier with a pre-computed kernel (cosine, Jaccard or Hamming kernel according to the embedding techniques) 
to return the most probable labels for each node. We report the average Macro-F1 and Micro-F1 scores from 5 repeated trials. 
A higher value of these metrics implies better performance.

\begin{table*}[t]\small 
\vspace{-4pt}
\caption{Node classification performance using kernel SVM, where the top-3 results for each dataset are highlighted with bold numbers
}
\label{tbl:classifkern}
\centering
\vspace{-4pt}
\rowcolors{4}{gray!25}{white}
\begin{tabular}{l|rrrr|rrrr}
\hline
\multirow{2}{*}{\textbf{Method}} & \multicolumn{4}{c|}{\textbf{Micro-F1 (\%)}}                                                 & \multicolumn{4}{c}{\textbf{Macro-F1 (\%)}}                                                  \\ \cline{2-9}  
                         & \textbf{Blog}    & \textbf{PPI}            & \textbf{Wiki}           & \textbf{DBLP}           
        & \textbf{Blog}    & \textbf{PPI}            & \textbf{Wiki}           & \textbf{DBLP}           
\\ \hline
$\cdot$DeepWalk                 & 39.60          & \textbf{17.24} & 46.05          & 83.46          
& 21.93          & \textbf{10.28} & 6.62           & 83.16          
\\
:Node2Vec                 & 37.95          & 16.04          & 50.32          & 93.25          
& 20.22          & 9.57           & 9.86           & 93.12          
\\
$\cdot$LINE                     & 35.49          & 15.01          & 48.22          & 86.83          
          & 16.60          & 8.70           & 8.47           & 86.54          
\\
$\cdot$VERSE                    & 39.61          & 15.90          & 41.39          & 92.79          
 & 22.85		  & 9.76           & 4.14           & 92.66          
\\ \hline
:GraRep                   & 36.21          & 5.83           & 56.22          & 91.41          
              & 16.91          & 1.52           & 12.14          & 91.25          
\\
:HOPE                     & 31.37          & 14.69          & \textbf{56.68} 		 & 91.47          
          & 11.74          & 8.13           & \textbf{13.30}  & 91.30          
\\
:NetMF                    & \textbf{40.04} & 15.03          & \textbf{57.62}          & \textbf{93.59} 
    & \textbf{23.43}  & 8.74  & \textbf{14.35} & \textbf{93.46} 
\\ \hline
$\cdot$INH-MF                   & \textbf{36.13} & 15.50 & 45.03 & \textbf{93.27} 
 			& \textbf{18.88} & 9.55  & 6.90  & \textbf{93.16} 
\\ 
:NetHash                  & 35.80          & 18.85          & 47.57          & 97.61          
          & 18.72          & 12.91          & 8.05           & 97.57          
\\
:NodeSketch               & \textbf{38.16} & \textbf{21.04} & \textbf{59.0}7          & \textbf{98.83} 
 & \textbf{21.84} & \textbf{15.55} & \textbf{16.31}          & \textbf{98.81} 
\\ \hline
$\cdot$HARP-DWalk               & 36.52 & 15.46 & 43.06         & 92.66 
& 19.56 & 9.04 & 5.59          & 92.53 
\\
$\cdot$HARP-LINE               & 30.27 & 12.67 & 42.79         & 88.07 
  & 13.06 & 6.25 & 5.38          & 87.84 
\\ \hline
DAOC               & 21.3 & 12.56 & 42.43 & 89.24          
 & 6.47 & 7.25 & 5.66          & 89.03 
\\
DAOR               & 33.05 & \textbf{19.07} & 53.24         & 87.86 
  & 17.25 & \textbf{13.94} & 15.97          & 87.64 
\\
\hline
\end{tabular}
\begin{flushleft}
\center
\footnotesize{
$\cdot$  the algorithm meta parameters are \emph{tuned} once for all datasets to maximize accuracy\\
:  the algorithm parameters are \emph{tuned for each dataset} to maximize accuracy
}
\end{flushleft}
\vspace{-4pt}
\end{table*}
Our method, \sys, shows competitive
results without requiring 
any tuning 
(unlike conventional embedding techniques, which require extensive tuning, as described above).
We also evaluated embeddings generated using DAOC without our proposed extension for multi-resolution clustering. The improvement of DAOR over DAOC verifies the effectiveness of our proposed extensions 
for graph embedding.  

\subsection{Link Prediction Task}

Link prediction is a typical graph analysis task that predicts potential (or missing) links between nodes in a graph. For this task, we use the same setting as in~\cite{Ou16}. Specifically, we randomly sample 20\% of the links out of the graph as test data, and use the rest of the graph for training. After learning the node embeddings based on the training graph, we predict the missing links by generating a ranked list of potential links. For each pair of nodes, we use the cosine, Jaccard or Hamming similarity (according to the embedding techniques) between their embedding vectors to generate the ranked list. As the number of possible pairs of nodes is too large, we randomly sample 0.1\% pairs of nodes for evaluation. We report the average precision@N and recall@N from 10 repeated trials.

\begin{table*}[htbp]\small 
\vspace{-4pt}
\caption{Link prediction performance, where the top-3 results for each dataset are highlighted with bold numbers
}
\label{res_link}
\centering
\vspace{-4pt}
\rowcolors{4}{gray!25}{white}
\begin{tabular}{l|rrrr|rrrr}
\hline
\multirow{2}{*}{\textbf{Method}} & \multicolumn{4}{c|}{\textbf{Precision@100}}                                   
& \multicolumn{4}{c|}{\textbf{Recall@100}}
\\ \cline{2-9} 
                         & \textbf{Blog}            & \textbf{PPI}             & \textbf{Wiki}            & \multicolumn{1}{r|}{\textbf{DBLP}}
         & \textbf{Blog}            & \textbf{PPI}             & \textbf{Wiki}            & \multicolumn{1}{r|}{\textbf{DBLP}}
\\ \hline
$\cdot$DeepWalk                 & 0.0200          & 0.0159          & 0.0090          & 0.0423                    
          & 0.0301          & 0.2227          & 0.0493          & 0.6749                    
\\
:Node2Vec                 & \textbf{0.0927} & 0.0137          & \textbf{0.0267} & 0.0321                    
& \textbf{0.1378} & 0.1958          & \textbf{0.1514} & 0.5174                    
\\
$\cdot$LINE                     & 0.0070          & 0.0073          & 0.0031          & 0.0392                    
          & 0.0103          & 0.0923          & 0.0167          & 0.6186                    
\\
$\cdot$VERSE                    & 0.0404          & \textbf{0.0206}   & 0.0212          & \textbf{0.0436}           
 & 0.0602          & \textbf{0.2723}          & 0.1118          & \textbf{0.6906}           
\\ \hline
:GraRep                   & 0.0014          & 0.0011          & 0.0054          & 0.0001                    
              & 0.0020          & 0.0118          & 0.0286          & 0.0011                   
\\
:HOPE                     & 0.0023          & 0.0073          & 0.0027          & 0.0248                    
          & 0.0035          & 0.0960          & 0.0149          & 0.4034                    
\\
:NetMF                    & 0.0175          & 0.0174		   & 0.0084          & 0.0218                    
               & 0.0266          & 0.2287 & 0.0474          & 0.3126                    
\\ \hline
$\cdot$INH-MF                   & 0.0158 & 0.0158 & 0.0084 & \textbf{0.0252}           
& 0.0227 & 0.2209 & 0.0454 & \textbf{0.4052}           
\\ 
:NetHash                  & 0.0015          & 0.0134          & 0.0020          & 0.0387                    
          & 0.0022          & 0.1899          & 0.0101          & 0.5958                    
\\
:NodeSketch               & \textbf{0.0729} & \textbf{0.0250} & \textbf{0.0176} & \textbf{0.0462}           
 & \textbf{0.1080} & \textbf{0.3331} & \textbf{0.0942} & \textbf{0.7595}           
\\ \hline
$\cdot$HARP-DWalk               & x & 0.0142 & 0.0101         & 0.0407 
& x & 0.1978 & 0.0536          & 0.6459 
\\
$\cdot$HARP-LINE               & x & 0.0026 & 0.0021 & 0.0309         
& x & 0.0331 & 0.0117          & 0.5029 
\\ \hline
DAOC               & 0.0001 & 0.0099 & 0.0059         & 0.0027 
& 0.0001 & 0.1301 & 0.0314          & 0.0444 
\\
DAOR               & \textbf{0.0958}  &  \textbf{0.0175}  & \textbf{0.0164}  & 0.0032            
& \textbf{0.1438}  & \textbf{0.2345}  & \textbf{0.0892}  & 0.0548          
\\ \hline
\end{tabular}
\begin{flushleft}
\center
\footnotesize{
$\cdot$  the algorithm meta parameters are \emph{tuned} once for all datasets to maximize accuracy\\
:  the algorithm parameters are \emph{tuned for each dataset} to maximize accuracy\\
x  the algorithm is crashed on coarsening small disconnected components
}
\end{flushleft}
\vspace{-4pt}
\end{table*}
Table~\ref{res_link} shows the results of the link prediction task. Our proposed method, \sys, is among the top-3 best-performing techniques being unsupervised and parameter-free. The impact of our proposed multi-resolution clustering is especially visible on this task, were \sys significantly outperforms DAOC.

\subsection{Robustness to the Metric Space}

Node embedding robustness to different metric spaces is shown in Table~\ref{tbl:metric} on the node classification task for our method \sys versus the two other best-performing methods from Table~\ref{tbl:classifkern} that technically can be evaluated with another metric space (NodeSketch is omitted because it uses a non-linear Hamming metric space, where cosine distance cannot be formally evaluated). For \emph{all} input graphs, \sys shows the best worst-case performance, i.e.,  \sys yields much more accurate results in its least accurate non-primary metric space than the other methods do. Moreover, Hamming distance is directly applicable to \sys without any preliminary binarization, unlike the algorithms operating in the cosine metric space.

\begin{table}[htbp]\small 
\vspace{-4pt}
\caption{Node embedding robustness to the metric space, where the native metric space for each algorithm is highlighted in bold}
\label{tbl:metric}
\centering
\vspace{-4pt}
\begin{tabular}{@{}ll|llll@{}}
\hline
\multirow{2}{*}{\textbf{Method}} & \multirow{2}{*}{\textbf{Metric}} & \multicolumn{4}{c}{\textbf{Micro-F1 (\%)}}                   \\ \cline{3-6}
                                 &                                  & \textbf{Blog} & \textbf{PPI} & \textbf{Wiki} & \textbf{DBLP} \\ \hline
\multirow{4}{*}{DAOR}            & \textbf{jaccard}                 & 33.05         & 19.07        & 53.24         & 87.86         \\
                                 & cosine                           & 30.41         & 13.62        & 47.20         & 87.42         \\
                                 & hamming                          & \textit{23.87}         & 14.13        & 45.52         & \textit{86.98}         \\
                                 & binham                          & 23.89         & \textit{10.25}        & \textit{41.36}         & 87.17         \\ \hline
\multirow{3}{*}{NetMF}           & \textbf{cosine}                  & 40.04         & 15.03        & 57.62         & 93.59         \\
                                 & hamming                          & 17.54         & 6.55         & 40.93         & 70.32         \\
                                 & binham                          & 19.82         & 7.05         & 42.85         & 74.91         \\ \hline
\multirow{3}{*}{HOPE}           & \textbf{cosine}                  & 31.37          & 14.69          & 56.68 		 & 91.47         \\
                                 & hamming                          & 17.02         & 5.95         & 40.89         & 59.94         \\
                                 & binham                          & 18.23         & 6.21         & 40.89         & 74.36         \\ \hline
\end{tabular}
\begin{flushleft}
\footnotesize{
binham  - Hamming metric applied after a binarization of each dimension using its median value
}
\end{flushleft}
\vspace{-4pt}
\end{table}

\subsection{Runtime Performance}

In this experiment, we investigate the efficiency of the graph embedding learning process. Our evaluation was performed using an open-source graph embeddings evaluation framework, GraphEmbEval\footref{ftn:gembev}. 
on a Linux Ubuntu 16.04.3 LTS server with an Intel Xeon CPU E5-2620 v4 @ 2.10GHz CPU (16 physical cores) and 132 GB RAM. The training and execution termination constraints for each algorithm were set to 64 GB of RAM and 240 hours CPU time (we terminate the process when either of those thresholds are met).

\begin{table}[htbp]\small 
\vspace{-4pt}
\caption{Node embedding learning time (in seconds), where the top 3 results for each dataset are highlighted in bold
}
\label{res_runtime_embs}
{
\vspace{-4pt}
\rowcolors{2}{gray!25}{white}
\begin{tabular}{l|rrrrr}\hline
\textbf{Method}           & \textbf{Blog} & \textbf{PPI}  & \textbf{Wiki} & \textbf{DBLP}  & \textbf{YouTube} \\ \hline 
DeepWalk          & 3375 & 1273 & 1369 & 4665  & 747060                                                        \\
Node2Vec          & 1073 & 383  & 1265 & 504   & -                                                             \\
LINE              & 2233 & 2153 & 1879 & 2508  & 29403                                                       \\
VERSE             & 1095 & 203  & 276  & 1096  & 245334                                                       \\ \hline  
GraRep            & 3364 & 323  & 422  & 10582 & -                                                            \\
HOPE              & 239  & 100  & 78   & 283   & 15517                                                        \\
NetMF             & 487  & 124  & 708  & 213   & -                                                         \\   \hline  
INH-MF            & 509  & 39   & 98   & 378   & -                                                              \\ 
NetHash           & 721  & 201  & 134  & 35    & 12708                                                         \\
NodeSketch        & \textbf{70}   & \textbf{8}    & \textbf{17}   & \textbf{8}     & \textbf{2439}               \\  \hline  
HARP-DWalk              & 1436   & 299    & 483   & 958     & 336200                                                            \\
HARP-LINE              & 1274   & 106    & 189   & 90     & 13951                                                  \\  \hline   
DAOC              & \textbf{7.9}   & \textbf{0.3}    & \textbf{1.8}   & \textbf{0.2}     & \textbf{4893}                                                            \\
DAOR              & \textbf{1.6}   & \textbf{0.2}    & \textbf{0.4}   & \textbf{0.2}     & \textbf{57.1}                                                            \\ \hline
\end{tabular}}
\begin{flushleft}
\footnotesize{
-  the algorithm was terminated by timeout
}
\end{flushleft}
\vspace{-4pt}
\end{table}

Table \ref{res_runtime_embs} shows the end-to-end embedding learning time. To discount the impact of the 
multi-threaded implementation of some of the methods, we dedicate a single logical CPU per each method implementation and report the total CPU time. 
Our method, \sys, is faster than existing state-of-the-art techniques by several orders of magnitude; it exhibits near-linear scaling when increasing the number of links in the graph.
\sys is also much more scalable than DAOC (on which \sys is built) due to its specific multi-scaling approach that boosts the number of nodes reaching consensus of the optimization function early on lower levels of the hierarchy. 
Moreover, unlike other graph embedding techniques, \sys execution time decreases when increasing the number of embedding dimensions, which is due to the early termination of the clustering as described in Section~\ref{subsec:clsbound}. 


%


\section{Conclusions}

In this paper, we presented a novel highly efficient and parameter-free graph embedding technique, \sys\hspace{-4pt}\footref{ftn:sys}, 
which produces metric-robust 
and interpretable embeddings without requiring any manual tuning. Compared to a dozen state-of-the-art graph embedding algorithms, \sys yields competitive results on diverse graph analysis tasks (node classification and link prediction), while being several orders of magnitude more efficient. 

In future work, we plan to recommend the minimal, maximal and optimal number of embedding dimensions, 
and conduct a comprehensive study on their quality and interpretability. Also, we plan to 
integrate further state-of-the-art community detection algorithms in addition to DAOC.


%
%

\bibliographystyle{IEEEtran}\balance
\bibliography{IEEEabrv,./daor}

\end{document}